\documentclass[10pt,conference,a4paper]{IEEEtran}
\usepackage[latin1]{inputenc}
\usepackage[cmex10]{amsmath}
\usepackage{amsfonts}
\usepackage{amssymb}
\usepackage{graphicx}
\usepackage{color} 
\usepackage{url} 
\usepackage{bm} 
\usepackage{enumerate} 
\usepackage{lhelp} 
\usepackage{csquotes} 

\title{Experience with Heuristics, Benchmarks \& Standards for Cylindrical Algebraic Decomposition}

\author{
	\IEEEauthorblockN{
	Matthew England\IEEEauthorrefmark{1}, 
	and
	James H. Davenport\IEEEauthorrefmark{2}, 
	}
	\IEEEauthorblockA{
	\IEEEauthorrefmark{1}Coventry University Faculty of Engineering, 
	Environment and Computing, Coventry, CV1 2JH, U.K.
	\\ Email: \texttt{Matthew.England@coventry.ac.uk}}
	\IEEEauthorblockA{
	\IEEEauthorrefmark{2}University of Bath Department of Computer Science, Bath, BA2 7AY, U.K.
	\\ Email: \texttt{J.H.Davenport@bath.ac.uk}}
}

\begin{document}

\maketitle

\begin{abstract} 
In the paper which inspired the \textsf{SC}$^2$ project, [E.~\'Abr\'aham, \textit{Building Bridges between Symbolic Computation and Satisfiability Checking}, Proc. ISSAC '15, pp. 1--6, ACM, 2015] the author identified the use of sophisticated heuristics as a technique that the Satisfiability Checking community excels in and from which it is likely the Symbolic Computation community could learn and prosper.  To start this learning process we summarise our experience with heuristic development for the computer algebra algorithm Cylindrical Algebraic Decomposition.  We also propose and discuss standards and benchmarks as another area where Symbolic Computation could prosper from Satisfiability Checking expertise, noting that these have been identified as initial actions for the new \textsf{SC}$^2$ community  in the CSA project, as described in [E.~\'Abr\'aham et al., \textit{\textsf{SC}$^2$: Satisfiability Checking meets Symbolic Computation (Project Paper)}, Intelligent Computer Mathematics (LNCS 9761), pp. 28--43, Springer, 2015].
\end{abstract}

\section{Introduction}
\label{sec:intro}

This article is inspired by the \textsf{SC}$^2$ project\footnote{http://www.sc-square.org/}, a new initiative to forge a joint community from the existing fields of \textsf{S}ymbolic \textsf{C}omputation and \textsf{S}atisfiability \textsf{C}hecking.  For further details on the project we refer the reader to:
\begin{itemize}
\item \cite{AAB+16a} which introduced the two fields, describes some of the challenges and opportunities from working together, and outlines planned project actions; and
\item \cite{Abraham2015} the accompanying paper to an invited talk at ISSAC 2015 which inspired the creation of the new project and community.
\end{itemize}
Within \cite{Abraham2015} the author outlines the strengths and weaknesses of the two communities, writing in the introduction:
\begin{displayquote}
``Symbolic Computation is strong in providing powerful procedures for sets (conjunctions) of arithmetic constraints, but it does not exploit the achievements in SMT solving for efficiently handling logical fragments, using heuristics and learning to speed-up the search for satisfying solutions.''
\end{displayquote}
By \emph{heuristic} the we mean a practical method to make a choice which is not guaranteed to be optimal.  Although Computer Algebra Systems prize correctness and exact solutions there is still much scope for the use of heuristics and statistical methods in symbolic computation algorithms: both for tuning how individual algorithm are run and for selecting a particular algorithm to use in the first place.  In regards to the latter point, we note that the \texttt{solve} procedures in Computer Algebra Systems are really meta-algorithms: algorithms to select specific procedures to use based on problem parameters.  Although the individual procedures are usually well documented within the scientific literature we are not aware of any publications describing these meta-algorithms.  

Another topic where Symbolic Computation might benefit from experience in Satisfiability Checking is standards and benchmarks.  Competitions based on these form an integral part of the Satisfiability Checking community, and may have contributed to the remarkable progress made in practical algorithms.  The lack of a comparable focus for the Symbolic Computation community was acknowledged in \cite{AAB+16a}.  However, recent experiments have suggested the benchmarks for non-linear real arithmetic are insufficient and the development of new standards and benchmarks for the joint community has been identified as a key \textsf{SC}$^2$ project action in \cite[Section 3.3]{AAB+16a}.  

In the present paper we outline our experience of these issues for a single Symbolic Computation algorithm, Cylindrical Algebraic Decomposition (CAD).  The aim of the paper is to instigate the learning process from the Satisfiability Checking community by illustrating the current use of heuristics, benchmarks and standards in (at least one area of) Symbolic Computation and posing some questions.  We start with a summary of the necessary background on CAD in Section \ref{sec:CAD}, then survey work with heuristics in CAD in Section \ref{sec:Heuristics} and our experience with standards and benchmarks in Section \ref{sec:SB}.  We finish with conclusions and questions in Section \ref{sec:Conc}.

\section{Cylindrical algebraic decomposition}
\label{sec:CAD}

\subsection{Definition}
\label{ssec:def}

\begin{figure}[b]
\centering
\includegraphics[width=0.45\columnwidth]{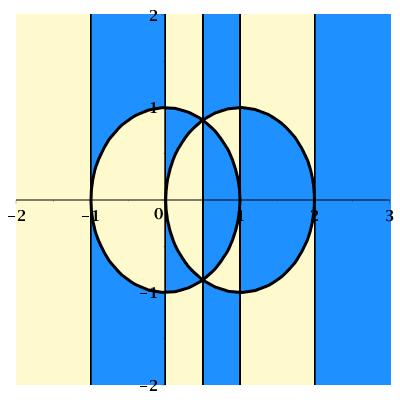}
\hspace*{0.05\columnwidth}
\includegraphics[width=0.45\columnwidth]{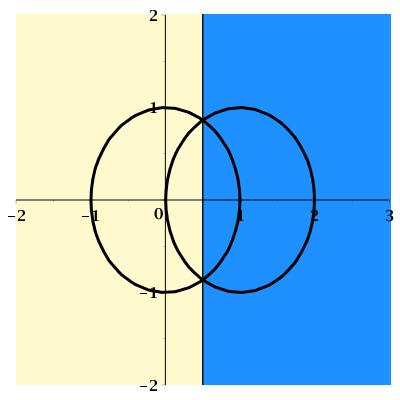}
\caption{Example visualising sign and truth-invariant CADs}
\label{fig:ex1}
\end{figure}

A \emph{Cylindrical Algebraic Decomposition} (CAD) is a \emph{decomposition} of $\mathbb{R}^n$ into \emph{cells} (connected subsets).  By \emph{algebraic} we mean semi-algebraic: i.e. each cell can be described with a finite sequence of polynomial constraints.  Finally, the cells are arranged \emph{cylindrically}, meaning the projections of any pair, with respect to the variable ordering in which the CAD was created, are either equal or disjoint.  We assume variables labelled according to their ordering (so the projections considered are $(x_1,\ldots,x_{\ell})\rightarrow(x_1,\ldots,x_k)$ for $k<\ell$) with the highest ordered variable present said to be the \emph{main variable}.  
Hence CADs can be represented in a tree like format branching on the semi-algebraic conditions involving increasing main variable, as in the example below (with the branching from right to left; all $\sqrt{}$ indicating the positive root; and the tuples on the left sample points of the cells).
\[
\begin{cases}
(-2,0) & \quad x<-1 \\ \\
\begin{cases} (-1,-1) & \quad y<0 \\ (-1,0) & \quad y=0 \\ (-1,1) & \quad 0<y \end{cases}  & \quad x=-1 \\ \\
\begin{cases} 
(0,-2) & \qquad y<-\sqrt{-{x}^{2}+1} \\ 
(0,-1) & \qquad y=-\sqrt {-{x}^{2}+1} \\ 
(0,0) & -\sqrt {-{x}^{2}+1}<y<\sqrt {-{x}^{2}+1} \\ 
(0,1) & \qquad y=+\sqrt {-{x}^{2}+1} \\ 
(0,2) & \qquad \sqrt {-{x}^{2}+1}<y 
\end{cases}
&-1<x<1 \\ \\
\begin{cases} (1,-1) & \quad y<0 \\ (1,0) & \quad y=0 \\ (1,1) & \quad 0<y \end{cases}  & \quad x=1  \\ \\
(2,0) & \quad 1<x
\end{cases}
\]
A CAD is produced to be invariant for input; originally \emph{sign-invariant} for a set of input polynomials (so on each cell each polynomial is positive, zero or negative).  The example above is a sign invariant CAD for the polynomial $x^2+y^2-1$ defining the unit circle.  More recently CADs have been produced \emph{truth-invariant} for input Boolean-valued polynomial formulae.  A sign-invariant CAD for the polynomials in a formula is also truth-invariant for the formula; but we can often achieve truth-invariance with far less cells.  For example suppose we need a CAD truth-invariant for the formula 
\[
x^2 + y^2 -1 = 0 \land (x-1)^2 + y^2 -1 = 0.
\]
A sign-invariant CAD would require 55 cells (with the full dimensional ones shown on the left of Figure \ref{fig:ex1}).  However, a truth-invariant CAD would need only 7 cells: 2 of which are full dimension (as on the right of Figure \ref{fig:ex1}) and 5 more to decompose the line $x=\tfrac{1}{2}$ at the points of intersection.

\subsection{Computation}
\label{ssec:Comp}

CAD construction usually involves two phases.  The first {\em projection}, applies operators recursively on polynomials, starting with the input.  Each time the operator produces a set with one less variable which together define the {\em projection polynomials}.  These are used in the second phase, {\em lifting}, to build CADs of increasing dimension.  First a CAD of $\mathbb{R}^1$ is built by splitting on the real roots of the univariate polynomials (those in $x_1$ only).  
Next, a CAD of $\mathbb{R}^2$ is built by repeating the process over each cell in $\mathbb{R}^1$ with the bivariate polynomials in ($x_1,x_2)$ 
evaluated at a sample point of the cell in $\mathbb{R}^1$; and the process is repeated until a CAD of $\mathbb{R}^n$ is produced.  We call the cells where a polynomial vanishes {\em sections} and those regions in-between {\em sectors}, which together form the {\em stack} over the cell.  
In each lift we extrapolate the conclusions drawn from working at a sample point to the whole cell requiring validity theorems for the projection operator used.

CAD cells are represented by at least a \emph{sample point} (as in the left of the example above), and an \emph{index}: a list of positive integers with each integer indicating the section or stack each variable is within (in reference to the ordered roots of the projection polynomials).  Some implementations will also encode the full algebraic description within each cell.

CAD was originally introduced by Collins for quantifier elimination (QE) in real closed fields \cite{ACM84I}.  Although CAD construction has complexity doubly exponential in the number of variables \cite{DH88}, applications range from parametric optimisation \cite{FPM05} and epidemic modelling \cite{BENW06}, to reasoning with multi-valued functions \cite{DBEW12} and the derivation of optimal numerical schemes \cite{EH14}. 
There have been many improvements to Collins' original approach most notably refinements to the projection operators \cite{McCallum1998} \cite{Brown2001a}, \cite{MPP16}; early termination of lifting \cite{CH91} \cite{WBDE14}; and symbolic-numeric schemes \cite{Strzebonski2006}, \cite{IYAY09}.  Some recent advances include dealing with multiple formulae \cite{BDEMW13}, \cite{BDEMW16}; local projection \cite{Brown2013}, \cite{Strzebonski2014a}; and decompositions via complex space \cite{CMXY09}, \cite{BCDEMW14}.  For a more detailed introduction to CAD see e.g. \cite{BDEMW16}.

\section{Heuristic use for CAD}
\label{sec:Heuristics}

\subsection{Choosing the variable ordering}
\label{ssec:ord}

The most well known choice required for CAD is that of the variable ordering, which the cylindricity is defined with respect to.  This determines the order in which variables are eliminated during projection and the subspaces through which CADs are built incrementally during lifting.  
When using CAD for QE we must project variables in the order they are quantified, but we are free to project the other variables in any order (and to change the order within quantifier blocks).  

\begin{figure}[b]
\centering
\includegraphics[width=0.48\columnwidth]{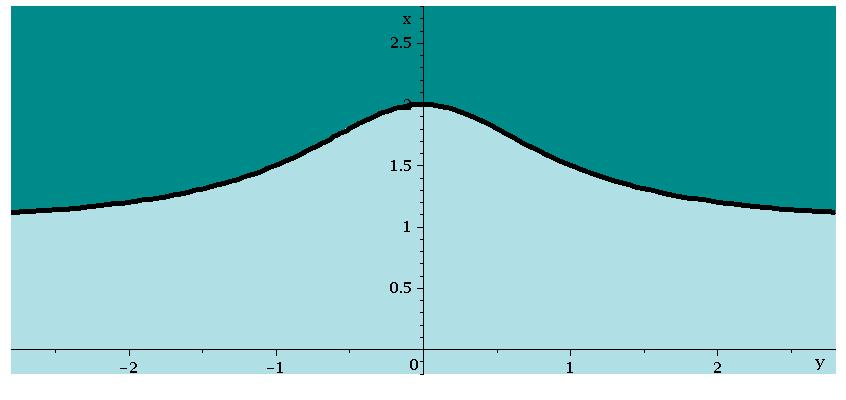}
\hspace*{0.01\columnwidth}
\includegraphics[width=0.48\columnwidth]{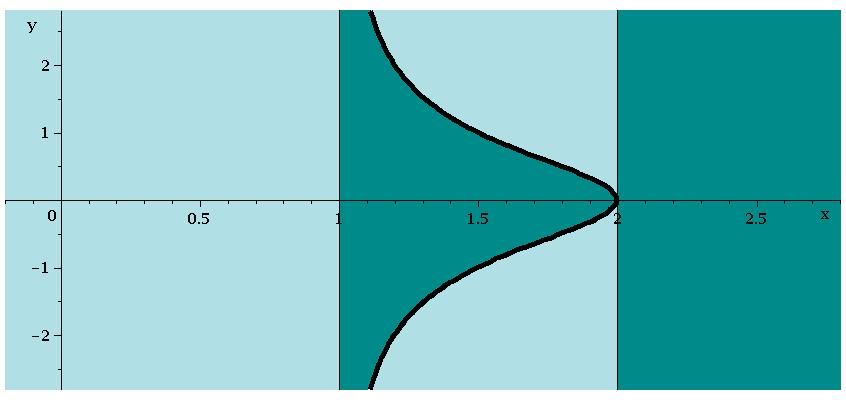}
\caption{CADs under different variable orderings}
\label{fig:ex2}
\end{figure}

The variable ordering used can have a great effect on the output produced.  For example, let $f:=(x-1)(y^2+1)-1$ and consider the minimal sign-invariant CAD in each variable ordering, as visualised in Figure \ref{fig:ex2}.  In each case we project down with the left figure projecting $x$ first and the right $y$.  In this toy example the \emph{``wrong''} choice more than doubles the number of cells, while numerous experiments have shown that for larger examples the choice can determine whether a problem is tractable (see for example the experimental results in \cite{BDEMW16}).
At the extreme end of this observation, \cite{BD07} defined a class of examples where changing variable ordering would change the number of cells required from constant to doubly exponential in the number of variables.

\noindent Several heuristics have been developed to choose the variable ordering, including:
\begin{description}
\item[\texttt{Brown}:] Use the following criteria, starting with the first and breaking ties with successive ones:
\begin{enumerate}[(1)]
\item Eliminate variable if lowest overall degree.
\item Eliminate variable if lowest (maximum) total degree in terms in which it occurs.
\item Eliminate variable if smallest number of terms contains it.
\end{enumerate}
Suggested by Brown in \cite[Section 5.2]{Brown2004}.
\item[\texttt{sotd}:] For all admissible orderings, calculate the projection set and choose the one with smallest \emph{sum of total degrees} for each of the monomials in each of the polynomials \cite{DSS04}.  Performs well but more costly than \texttt{Brown}.  
A greedy alternative is to allocate one variable of the ordering at a time by projecting each unallocated variable and choosing the one which increases the \texttt{sotd} least.
\item[\texttt{ndrr}:] As with \texttt{sotd} construct the full projection set and choose the ordering whose set has the least \emph{number of distinct real roots} of the univariate polynomials within \cite{BDEW13}. 
Even more costly than \texttt{sotd}, but sensitive to the real geometry and shown to assist with examples where the \texttt{sotd} heuristic failed.
\end{description}
The papers cited each contain experimental results demonstrating their worth.  In \cite{HEWDP14} the results of an experiment comparing the 3 on a data set of over 7000 examples were reported.  The experiments showed \texttt{Brown} selecting the best ordering (as measured by lowest cell count) more often than the others. However a key finding was that there were substantial subsets of examples for which each heuristic did best. Further, when calculating the saving made by the heuristics (compared to the average cell count of the different orderings) the authors of \cite{HEWDP14} found that \texttt{sotd} actually made a greater saving for quantified problems on average (i.e. while \texttt{Brown} was superior on more examples, \texttt{sotd} was superior on examples with greater savings on offer).
Together, this meant that recommending one heuristic at the expense of all others was not possible.  

If the minimal cell count is a priority then one further approach, suggested in \cite{WEBD14} is to compute the full dimensional cells for each possible ordering and pick the ordering with the minimum to derive a full CAD.  Computing full dimensional cells avoids any work with algebraic numbers and so is not as costly as may be thought, although it does require more computation than \texttt{ndrr}.  It was noted in \cite{WEBD14} that the full-dimensional cell calculations could be done in parallel with the first to finish extended to a full CAD and the rest discarded.

\subsection{Choices with equational constraints}
\label{ssec:EC}

\subsubsection{Equational constraints}

There are several ways in which we can modify CAD constriction to achieve truth-invariance (rather than sign-invariance) including refining sign-invariant CAD and truncating lifting once the truth-value is determined.  A particularly fruitful approach is to take advantage of the Booloean structure of a formula through the identification of \emph{Equational Constraints} (ECs): polynomial equations logically implied by a formula.
Reduced projection operators (using a subset of the usual polynomials) have been proven valid for use when an EC is present with corresponding main variable \cite{McCallum1999b, McCallum2001}.  Results in \cite{EBD15, ED16} suggest that the double exponent in the complexity bound decreases by 1 for each EC used, although this is restricted to primitive ECs meaning the classic lower-bound examples of \cite{DH88, BD07} are not violated \cite{DE16}.

These reduced operators can only use one EC for each projection, so when there are multiple we must make a \emph{designation}.  Note that ECs need not appear explicitly as atoms (formula with no logical connectives) in the input formula, but could instead be implicit.  For example, the resultant of any two ECs in the same main variable is itself an EC (not containing that variable) \cite{McCallum2001}.  Propagating ECs in this way allows for the maximal use of the reduced operators.  However, it can require multiple choices of which EC to designate at each projection.  Section 4 of \cite{EBD15} described such an example where the \emph{wrong} designation could make add tens of thousands to the cell count of the final output, making it more than 15 times bigger.

\subsubsection{Making the designation}

In \cite{BDEW13} the authors experimented in using the \texttt{sotd} and \texttt{ndrr} measures on this question (the \texttt{Brown} heuristic was not applicable since it acted only on the input polynomials).  In general they were useful in identifying the optimal designation, although both could be misled.  As described above these heuristics essentially complete the projection stage of the algorithm for each ordering, which although minimal in comparison to the lifting stage, is likely far more computation than would normally be undertaken by a heuristic.  This becomes an issue when the number of choices grows.  Further, for these experiments a fixed variable ordering was used, and the question of addressing the two choices together (when the number of possibilities multiplies) has not been addressed.  

\subsubsection{Designation in TTICAD}

In \cite{BDEMW13, BDEMW16} a truth-table invariant CAD (TTICAD) is defined as a CAD on whose cells the truth-table for a set of formulae is invariant.  A new operator was presented which takes advantage of ECs present in the separate formulae (with \cite{BDEMW13} developing the theory in the case where all had an EC and \cite{BDEMW16} extending to the general case).  The operator essentially recognises when to consider the interaction of polynomials from different formulae.  If an individual formula has multiple ECs then, as above, we must choose just one to designate for each projection.

\subsection{Choices for CAD by Regular Chains}
\label{ssec:RC}

Recently an alternative CAD computational scheme has been proposed where, instead of projection and lifting, we: first cylindrically decompose complex space according to whether polynomials are zero or not using the theory of triangular decomposition by regular chains; and then refine to a CAD of real space.  This was first proposed in \cite{CMXY09} with an incremental version described in \cite{CM14b} and an extension to TTICAD in \cite{BCDEMW14}.  All versions are implemented within the \texttt{RegularChains} Library\footnote{www.regularchains.org} with a summary in \cite{CM14a}.


Most of the heuristics outlined earlier in this section are not directly applicable to the CAD by Regular Chains computation scheme (as there is no \emph{``cheap''} projection phase to derive information from).  We outline some of the new heuristics developed for the choices this scheme involves.

\subsubsection{Variable order in TTICAD by Regular Chains}

This problem was considered in \cite{EBDW14}.  Two existing heuristics were compared: that of Brown introduced in Section \ref{ssec:ord} and another, denoted Triangular, already in use for other algorithms in the \texttt{RegularChains} Library.  Triangular chooses first the variable with lowest degree occurring in the input; then breaks ties by choosing variables for which leading coefficients have lowest total degree; and finally sum of degrees in input.  In addition the heuristics \texttt{sotd} and \texttt{ndrr} discussed above were used (even though the sets of projection polynomials built were not explicitly used later).  The experiments found \texttt{sotd} to make the best choices, but due to its higher costs the Triangular heuristic was the most efficient choice overall.  However, as with the experiments discussed in Section \ref{ssec:ord}, the example set could be subdivided into groups where different heuristics were dominant. Further experimentation and illustrative examples in \cite{EBDW14} led to the development of a new heuristic (composed from parts of the others) tailored to the variable ordering choice for TTICAD by Regular Chains \cite{BCDEMW14}.

\subsubsection{Constraint order in TTICAD by Regular Chains} 

The latest CAD algorithm within the \texttt{RegularChains} Library \cite{CM14b} processes constraints incrementally when building the complex cylindrical decomposition and thus are sensitive to the order in which constraints are considered.  Further, in the case of TTICAD we have the extra question of what order to consider the formulae in.  These issues were studied in \cite{EBCDMW14} which considered the following example.

\begin{figure}[t]
\centering
\includegraphics[width=0.49\columnwidth]{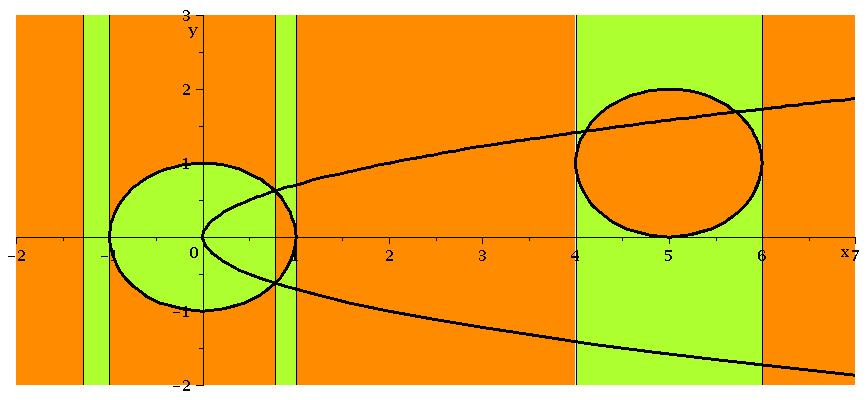}
\includegraphics[width=0.49\columnwidth]{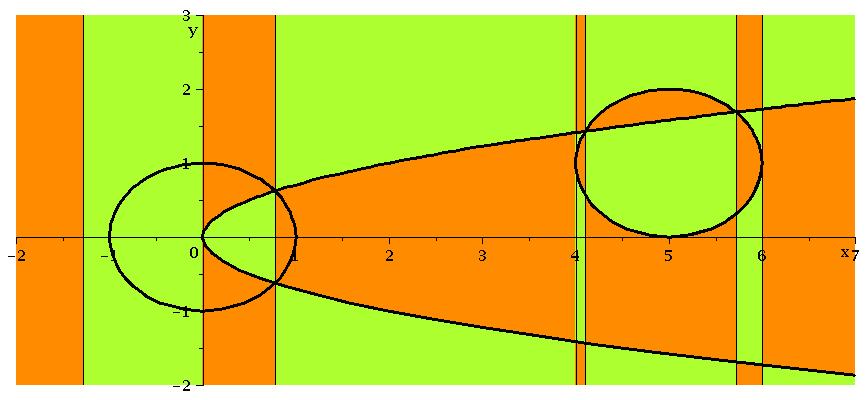}
\includegraphics[width=0.49\columnwidth]{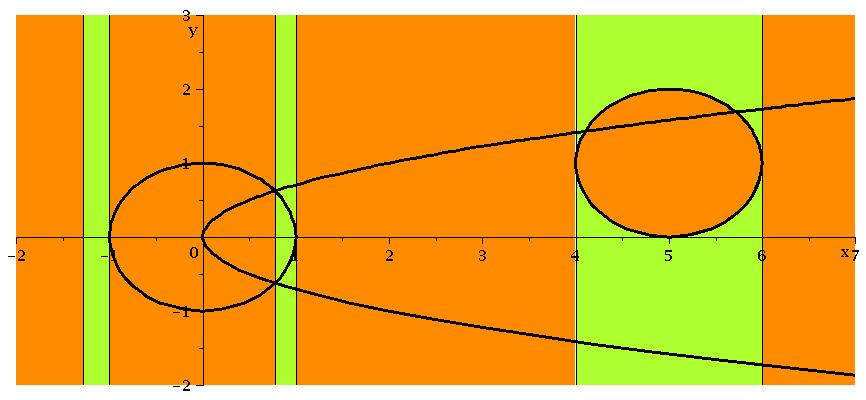}
\includegraphics[width=0.49\columnwidth]{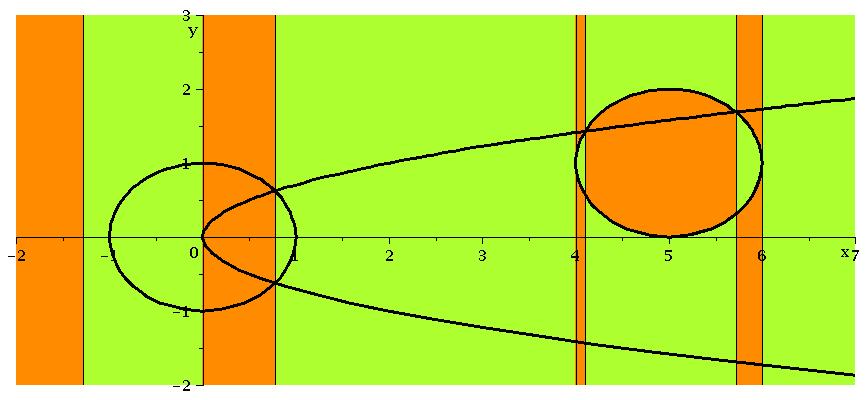}
\caption{Visualisations of the four TTICADs which can be built using the Regular Chains Library for the example in this section.  The figures on the top have $\phi_1 \rightarrow \phi_2$ and those on the bottom $\phi_2 \rightarrow \phi_1$.  The figures on the left have $f_1 \rightarrow f_2$ and those on the right $f_2 \rightarrow f_1$.
}
\label{fig:RCex}
\end{figure}

Assume the ordering $x \prec y$ and consider
\begin{align*}
f_{1} &:= x^2+y^2-1, \qquad  \\
f_{2} &:= 2y^2-x, \\
f_{3} &:= (x-5)^2+(y-1)^2-1,  
\\
\phi_1 &:= f_1=0 \land f_2 = 0, \qquad \\
\phi_2 &:= f_3 = 0.
\end{align*}
The polynomials are graphed within the plots of Figure \ref{fig:RCex}.  If we want to study the truth of $\phi_1$ and $\phi_2$ (or say a parent formula $\phi_1 \lor \phi_2$) we need a TTICAD to take advantage of the ECs.   There are two possible orders for the formulae and two possible to consider the constraints within $\phi_1$.  Hence 4 possible ways we could calculate a TTICAD by Regular Chains.
Below we show how many cells are produced by proceeding in the orders indicated, with the two dimensional cells shown in Figure \ref{fig:RCex}.
\begin{itemize}
\item \textbf{$\bm{\phi_1 \rightarrow \phi_2}$ and $\bm{f_1 \rightarrow f_2}$:} 
37 cells.
\item \textbf{$\bm{\phi_1 \rightarrow \phi_2}$ and $\bm{f_2 \rightarrow f_1}$:} 
81 cells.
\item \textbf{$\bm{\phi_2 \rightarrow \phi_1}$ and $\bm{f_1 \rightarrow f_2}$:} 
25 cells.
\item \textbf{$\bm{\phi_2 \rightarrow \phi_1}$ and $\bm{f_2 \rightarrow f_1}$:}
43 cells.
\end{itemize}

No previously discussed heuristic was applicable to this problem.  For choosing which EC to process first in a given formula an argument could be made for measuring a set of polynomials shown to be rendered sign-invariant by the algorithm (leading to Heuristic 1 in \cite{EBCDMW14}).  The only heuristic derived for the other choices was to measure the sum of degrees of the polynomials in the complex cylindrical decomposition created.  As with the \texttt{ndrr} and full dimensional cells heuristics above; this requires more computation than is ideal (although it is the real root refinement that makes up the bulk of the CAD by Regular Chains computation time).

\subsection{Gr\"obner Basis preconditioning}
\label{ssec:GB}

A \emph{Gr\"obner Basis} $G$ is a particular generating set of an ideal $I$ defined with respect to a monomial ordering \cite{Buchberger2006}.  One definition is that the ideal generated by the leading terms of $I$ is generated by the leading terms of $G$.  Gr\"obner Bases (GB) are used extensively to study ideals and the polynomials that define them as they allow properties such as dimension and number of zeros to be easily deduced.  Although like CAD the calculation of GB is doubly exponential in the worst case \cite{MM82}, GB computation is now mostly trivial for any problem on which CAD construction is tractable.  

It was first observed in \cite{BH91} that replacing a conjunction of polynomial equalities in a CAD problem by their GB (logically equivalent) could be useful for the CAD computation.  
Of the ten test problems studied: 6 were improved by the GB preconditioning (speed-up varying from 2- to 1700-fold); 1 problem resulted in a 10-fold slow-down; 1 timed out when GB preconditioning was applied, but would complete without; and 2 were intractable both for CAD construction alone and the GB preconditioning step.  
The problem was revisited in \cite{WBD12_GB}.  As expected, there had been a big decrease in the computation timings, especially for the GB. However, it was still the case that 2 of the problems were hindered by GB preconditioning.

The key conclusion is that GB preconditioning will on average benefit CAD (sometimes very significantly) but could on occasion  hinder it (to the point of making a tractable CAD problem intractable).  We are yet to understand why this occurs, but the authors of \cite{WBD12_GB} did develop a metric to predict when it will.  
They defined the \emph{Total Number of Indeterminates} (\texttt{TNoI}) of a set of polynomials $A$ as
\begin{equation*}
\label{eq:TNoI}
\texttt{TNoI}(A) = \sum_{a \in A} \texttt{NoI}(a)
\end{equation*}
where \texttt{NoI}$(a)$ is the number of indeterminates in a polynomial $a$.  The heuristic is to build a CAD for the preconditioned polynomials only if the \texttt{TNoI} decreased.  For most of their test problems the heuristic made the correct choice, but there were examples to the contrary.

\subsection{Use of machine learning}
\label{ssec:ML}

Finally, we note recent experiments using machine learning, specifically support vector machines (see for example \cite{STV04}), to make choices for CAD construction:
\begin{itemize}
\item In \cite{HEWDPB14} the authors used an SVM to choose between the three heuristics for CAD variable ordering outlined in Section \ref{ssec:ord}.  Simple problem features were selected (e.g. degrees, proportion of monomials containing each variable) and parameter optimisation was applied to maximise Matthews'  Correlation Coefficient \cite{Matthews1975}.  Over 7000 examples were studied, and over the 1721 reserved for the test set the machine learned choice was found to outperform each heuristic individually on average.
\item  In \cite{HEDP16} the authors used an SVM to predict when it will be useful to precondition a CAD problem with GB (see Section \ref{ssec:GB}).  The features used were from both the original input and the GB: so the study was answering the question \emph{should we use this GB} rather than \emph{should we compute it} (relevant since the GB computation was trivial for the problem set involved).  The machine learned choice outperformed both using GB universally, and the human defined \texttt{TNoI} heuristic.
\end{itemize}
We also note that a recent paper \cite{KIMA16} applied a support vector machine (seeded with the problem features from \cite{HEWDPB14}) to suggest the order in which QE should be performed on sub-formulae of a non-prenex formula. Experimental results on more than 2,000 non-trivial examples showed that machine learning was doing better than the human derived heuristics, following appropriate parameter optimisation.

\section{Standards and Benchmarks}
\label{sec:SB}

\subsection{History of benchmarking in computer algebra}
\label{ssec:SBHistory}

The Computer Algebra community has occasionally recognised the importance of benchmarks. The PoSSo and FRISCO projects aimed to do this for polynomials systems and symbolic-numeric problems respectively in the 1990s.  PoSSO, with which the second author was involved, collected a series of benchmark examples for GB, and a descendant of these can still be found online\footnote{\url{http://www-sop.inria.fr/saga/POL/}}.  However, this does not appear to be maintained; and the polynomials are not stored in a machine-readable form.  Polynomials from this list still crop up in various papers, but there is no systematic reference, and it is not clear whether people are really referring to the same example. Several of the examples are families, which is good but means that a benchmark has to contain specific instances.
The PoSSo project did its best to do ``level playing field'' comparisons, but at the time different implementations ran on different hardware / operating systems meaning this was not really achievable. The environment is much simpler these days, and it would be feasible to organise true contests.

The topic of benchmarking in computer algebra has most recently been taken up by the SymbolicData Project\footnote{www.symbolicdata.org} \cite{GNJ14} which is beginning to build a database of examples in XML format (although currently not with any suitable for CAD).  The software described in \cite{HL15} was built to translate problems in that database into executable code for various computer algebra systems.  The authors of \cite{HL15} discuss the peculiarities of computer algebra that make benchmarking particularly difficult including the fact that results of computations need not be unique and that the evaluation of the correctness of an output may not be trivial (or may be the subject of research itself). 

A final point to note is that while SAT / SMT-solvers have only a few clear possible answers (e.g. sat, unsat, unknown) in computer algebra there is also the quality of the result to consider (e.g., size of quantifier-free formula produced).  With CAD the output size is usually correlated to computation time, but this is not always the case with other algorithms.

\subsection{The present authors' recent experience with CAD}
\label{ssec:experience}

In our work we have taken various approaches to experimenting with CAD:
\begin{enumerate}[1]
\item (a) Test new results on examples previously used to evaluate CAD algorithms.
\end{enumerate}
For example, the experiments in \cite{WBD12_GB} started with the 10 examples in \cite{BH91} while \cite{DBEW12} and \cite{WDEB13} focussed on classic examples from \cite{Kahan87} and \cite{Davenport1986}.  The latter were derived from applications of CAD while the former seem to be a collection of geometric problems invented by the authors.  Other  papers that have contributed such test problems include \cite{Arnon1988}, \cite{McCallum1988} \cite{Brown2001b}, \cite {DSS04}.  We wonder whether historic repetition within the literature is alone a strong enough reason to be benchmark?

In \cite{WBD12_EX, Wilson2013} an attempt was made to gather together all those test examples in the literature for CAD, along with references of their first appearance in the literature and encodings for some computer algebra systems.

\begin{enumerate}[1]
\item (b) Supplement existing examples with modified versions suitable for demonstrating the feature in question.
\end{enumerate}
In \cite{BDEMW13, BDEMW16} the new TTICAD algorithm that was the subject of the paper offered an improvement on the state of the art for examples consisting of multiple formulae; or a single formulae in disjunctive normal form.  Such examples had not been the topic of any CAD papers before and no existing examples were capable of demonstrating the savings on offer.  The experiments produced in these papers were made of up two sets: 
\begin{itemize}
\item Formulae produced to describe the branch cuts of multivalued functions in a proposed simplification formula \cite{EBDW14}, with CAD to be applied so that the complex domain could be decomposed into regions where the functions were univariate, and thus the formula applicable or not.
\item Formulae produced by adapting the logical connectors in previous examples from the literature in \cite{WBD12_EX} so that conjunctions became disjunctions.
\end{itemize}
Clearly the former set is of great interest as they represent a real example for the algorithms; but they all conform to a single structure and so are arguably too uniform to alone draw broad conclusions from.  The second set were produced to be somehow close to the \emph{accepted} test examples of the literature, but whether this is any better than inventing a new examples from scratch is debatable.

\begin{enumerate}[2]
\item Derive new sets of random examples
\end{enumerate}

A recent experiment using machine learning \cite{HEWDPB14, HEDP16} (see Section \ref{ssec:ML}) exposed a shortcoming in the above techniques.  To train the SVMs hundreds of examples are required (with hundreds more then needed for validation and testing). The dataset from the literature in \cite{WBD12_EX} contained not nearly enough examples and while the datasets discussed in the next section were sufficient for the first experiment in \cite{HEWDPB14} they proved too uniform for \cite{HEDP16}.  We were left with no choice but to generate large quantities of new examples, which we did using the random polynomial generator in \textsc{Maple}.  We had applied this technique also in \cite{EBCDMW14, EBDW14} receiving positive feedback from reviewers for the technique; but the initial reviews of \cite{HEDP16} were all negative on the use of random data.  It seems the appropriateness of this technique varies with the community (conference) in question.  We opine that had we used data from the example bank of MetiTarski examples discussed in the next section then reviewers may have praised the focus on examples from a real application; even though MetiTarski themselves derive examples for benchmarks using random polynomials.

\subsection{Sources of large benchmarks sets}
\label{ssec:otherBM}

We note some other sources of large sets of benchmark problems that represent real applications of CAD:
\begin{itemize}
\item MetiTarski\footnote{\url{https://www.cl.cam.ac.uk/~lp15/papers/Arith}} \cite{AP10, Paulson2012} is an automatic theorem prover designed to prove theorems involving real-valued special functions (such as log, exp, sin, cos and sqrt).  In general this theory is undeciadable but MetiTarski is able to solve many problems by applying real polynomial bounds and then using real quantifier elimination tools like CAD.  Applications of MetiTarski in turn derive examples for CAD.  
\item The NRA (non-linear real arithmetic) category of the SMT-LIB library\footnote{\url{http://smtlib.cs.uiowa.edu/}} which according to \cite{JdM12} consists mostly of problems originating from
attempts to prove termination of term-rewrite systems. 
\end{itemize}

These two data sets where included in the nlsat Benchmark Set\footnote{\url{http://cs.nyu.edu/~dejan/nonlinear/}} produced to evaluate the work in \cite{JdM12}.  This also included verification conditions from the Keymaera \cite{PQR09} and parametrized generalizations of the problem from \cite{Hong1991}.  Together this dataset had many thousands of problems.  However, we note that the problems come from a small number of classes and may have some hidden uniformity.  

As mentioned above, the nlsat dataset was unsuitable to use for our machine learning experiment in \cite{HEDP16}.  Every single problem within that had more than one equality was aided by GB preconditioning, in fact a great many simply had a GB containing only 1 indicating the problem had no solution.  Previous experiments on small example sets suggested GB preconditioning sometimes harms CAD computation and this was verified by analysis of a large randomly generated dataset in \cite{HEDP16}.  Thus while the nlsat dataset is an excellent starting point it needs to be expanded to be less uniform.

We finish this section by noting one possible source of examples for the future.
\begin{itemize}
\item The Todai Robot Project\footnote{\url{http://21robot.org}} \cite{MIAA14} is a Japanese AI project that aims to have an artificial intelligence pass the entrance examination for the University of Tokyo by 2021. A majority of questions on the Mathematics exam can be resolved by real quantifier elimination with a variety of techniques employed \cite{IMAA14}.  A key difficulty is that the natural language processing of the question derives a formula of far greater complexity than the human derived equivalent.  This process derives a large bank of examples of CAD problems, as discussed in \cite{KIMA16} for example.  The authors of \cite{KIMA16} told us there are plans to make this data set public.
\end{itemize} 

\section{Conclusions and questions}
\label{sec:Conc}

After surveying the work in Section \ref{sec:Heuristics} we see that several approaches to the creation of heuristics have been taken: ranging from human identified algebraic features, justified by mathematical arguments and observations to different extents; to machine learned choices using a support vector machine.  We are interested to hear how these compare with the heuristics used in SAT-solvers and what lessons can be learned. 

We can identify at least two areas where CAD is in need of further heuristic development.
\begin{itemize}
\item \textbf{How best to take decisions in tandem:}  The work surveyed all considered choices to be made for CAD \emph{in isolation} (assuming other choices had already been fixed).  Of course, in reality this may not be the case.  We must decide which decisions to prioritise; how heuristics can be combined; and how the combinatorial blow-up of decisions can be contained.  Does the SAT-solving community have experience in similar issues?
\end{itemize}
There are many implementations of CAD including: the dedicated command line program
{\textsc Qepcad} \cite{Brown2003a};
{\textsc Mathematica} \cite{Strzebonski2006, Strzebonski2010}, where CAD is not available directly but is used as a subroutine for quantifier elimination; 
the \texttt{Redlog} package for \textsc{Reduce} \cite{SS03};
and 3 different \textsc{Maple} libraries - the \texttt{RegularChains} Library (see Section \ref{ssec:RC}; \texttt{SyNRAC} \cite{YA06} (now part of the Todai Robot project) and our own \texttt{ProjectionCAD} module \cite{EWBD14}.
\begin{itemize}
\item \textbf{How to choose between different implementations?}  
Each implementation includes unique pieces of theory and features and excels on different examples.  Ideally, we would have a single implementation which encompasses all recent advances.  A more manageable step may be an overarching \textsc{Maple} command to choose between the 3 packages there.  SMT-solvers are designed to use a variety of different theory solvers and how they choose between these may offer valuable lessons here.
\end{itemize}

Surveying Section \ref{sec:SB} raises a number of questions about how benchmark sets should be produced:
\begin{itemize}
\item How best to generate large numbers of examples which are not internally uniform?
\item How important is it that the benchmarks come from current applications?
\item How important is it that the benchmarks have historically been used in the literature?
\item Are randomly generated examples a fairer way to evaluate the software, or irrelevant as too far removed from applications?
\end{itemize}
The SAT / SMT community posses a unified, large and growing set of benchmarks in the SMT-LIB library and so we may be able to extrapolate lessons from this.  However, as noted above, this library may be too uniform \cite{HEDP16}, and comments from the anonymous referees suggest that this and other critics are already a topic of discussion in the SMT community.

\subsection*{Acknowledgements}

Thanks to our collaborators on the work surveyed here: 
Russell Bradford, 
James Bridge, 
Changbo Chen, 
Zongyan Huang, 
Scott McCallum, 
Marc Moreno Maza, 
Lawrence Paulson, and 
David Wilson.
Thanks also to the anonymous referees for their comments which improved the paper.

Most of the work surveyed here was supported by EPSRC grant EP/J003247/1.  The authors are now supported by EU H2020-FETOPEN-2016-2017-CSA project $\mathcal{SC}^2$ (712689).

\bibliographystyle{plain}
\bibliography{CAD}

\end{document}